\documentclass[aps,a4paper,prl,twocolumn,floatfix,showpacs]{revtex4} 
\usepackage{graphics}
\usepackage{graphicx}
\usepackage{amssymb}
\usepackage{amsfonts}
\usepackage{amsmath}
\usepackage{bm}
\usepackage{epsf}
\usepackage{version}
\usepackage[usenames]{color}

\newcommand \be  {\begin{equation}}
\newcommand \ee  {\end{equation}}
\newcommand \bea {\begin{eqnarray}}
\newcommand \eea {\end{eqnarray}}

\newcommand \bd  {\begin{details}}
\newcommand \ed  {\end{details}}
\begin{document}

\title{Quantum oscillations in antiferromagnetic conductors  
with small carrier pockets}
\author{Revaz Ramazashvili}
\affiliation{Laboratoire de Physique Th\'eorique  -- IRSAMC, CNRS 
and Universit\'e de Toulouse, UPS, F-31062 Toulouse, France}

\date{\today}

\excludeversion{details}


\begin{abstract}
I study magnetic quantum oscillations in antiferromagnetic conductors  
with small carrier pock\-ets and show, that combining the oscillation data 
with symmetry arguments and with the knowledge of the possible positions 
of the band extrema may allow to greatly constrain or even uniquely 
determine the location of a detected carrier pock\-et in the Brillouin zone. 
\end{abstract}


\maketitle

For over fifty years, magnetic quantum oscillations have been used as a direct 
and precise probe of the Fermi surface physics in metals \cite{shoenberg}. 
The scope of the quantum oscillation experiments has been ever expanding 
to new materials such as layered and chain compounds, 
mag\-ne\-ti\-cal\-ly ordered metals and superconductors. 

Recently, quantum oscillations were successfully observed in 
YBa$_2$Cu$_3$O$_{6+x}$ (YBCO) cuprate superconductors 
\cite{doiron,yelland,bangura,jaudet,sebastian_1,audouard}, 
prominent members of the family of doped antiferromagnetic insulators. 
In the underdoped region of the phase diagram, well-defined charged 
quasiparticles with a small-pock\-et Fermi surface were the key finding, 
whose further systematic study has only begun.  

The small size of the carrier pock\-ets points to an electron ordering and 
a concomitant Fermi surface reconnection -- and several types of order,  
including the {\em ortho-}II chain structure \cite{cafimov}, stripe-like spin 
density wave \cite{millis,harrison} and field-induced antiferromagnetism 
\cite{chen} were evoked to account for the observed area of the 
pockets. Distinguishing between these possibilities purely theoretically 
appears problematic: to reach agreement with quantum oscillation data, 
band structure calculations often require rigid shifts in the relative positions 
of the bands \cite{cafimov} and fitting renormalization factors \cite{millis}. 
These {\em ad hoc} adjustments may become substantial for small carrier 
pockets, let alone the unidentified nature of the electron order likely affecting 
the band structure in an unknown way. 
Given that probing the YBCO Fermi surface by angle-resolved photoemission 
remains a challenge, it is desirable to distinguish between the various ordering 
scenarios by means of only the quantum oscillations. Which invites a question, 
re\-le\-vant far beyond the physics of the cuprates: how do various types of order 
ma\-ni\-fest themselves in the quantum oscillations -- and how much can one 
possibly learn about a given type of order from a quantum oscillation 
measurement alone? 
\begin{figure}[h]
 \hspace{3cm}
 \epsfxsize=8cm
 \includegraphics[width=2.5in]{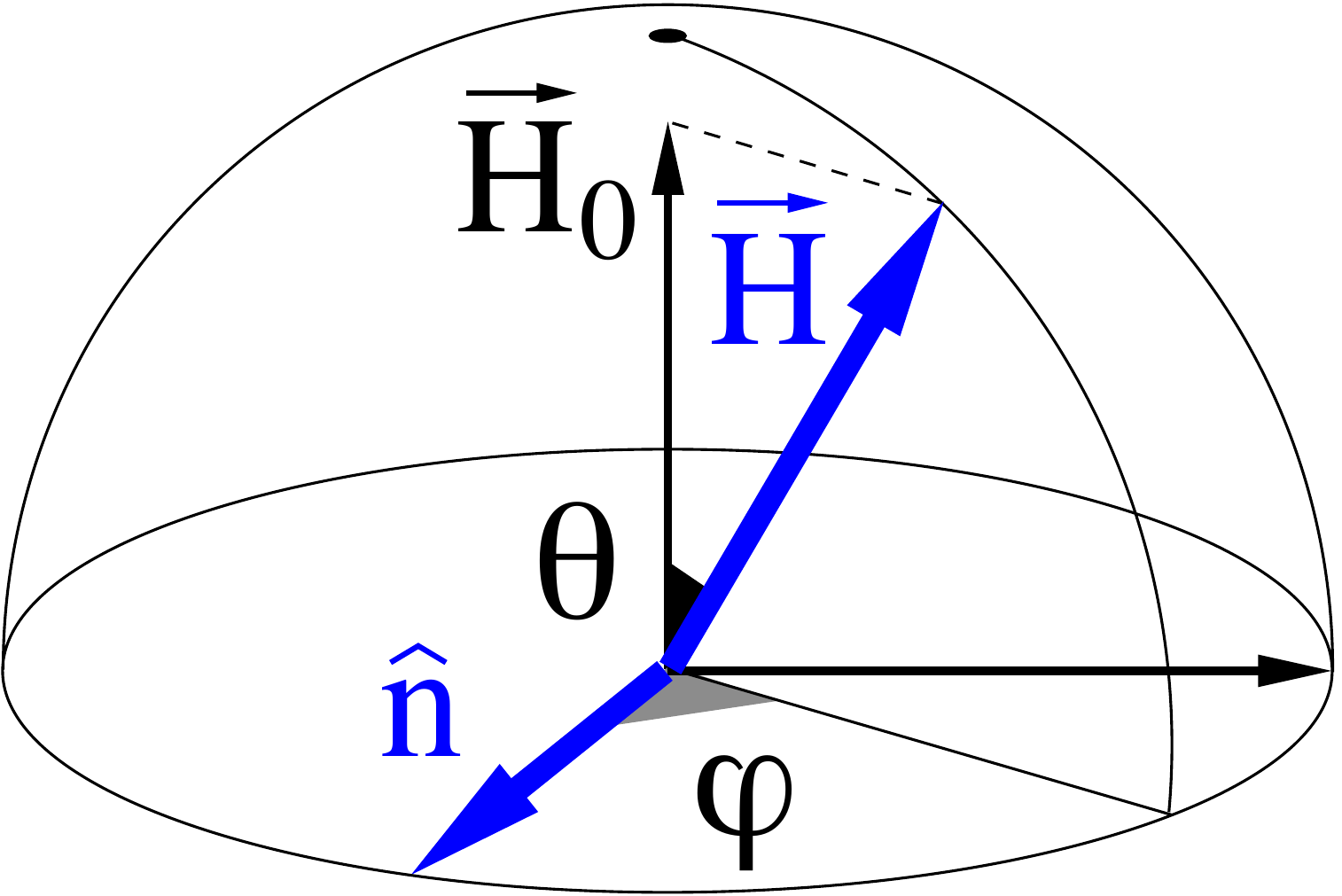}
 \vspace{10pt}
\caption{(color online). 
The staggered magnetization ${\bf n}$, pointing along the conducting plane, 
the magnetic field ${\bf H}$ and its normal component ${\bf H}_0$ with respect 
to the conducting plane. The orientation of the field is defined by the inclination 
angle $\theta$ and by the azimuthal angle $\varphi$, as shown in the figure.
 } 
\label{fig:real_space}
\end{figure}

An important step in this direction has been undertaken recently by Kabanov 
and Alexandrov \cite{kabanov}, who stu\-di\-ed the effect of the Zeeman 
splitting on the quantum oscillations in a weakly-doped two-dimensional 
insulator of square symmetry with the N\'eel antiferromagnetic order. 
The authors studied the reduction factor \mbox{$R_s$}, modulating the $n$-th 
harmonic amplitude due to interference of the contributions from 
the two Zeeman-split branches of the spectrum \cite{shoenberg} 
\be
\label{eq:amplitude} 
R_s = \cos \left[
\pi n \frac{\delta\mathcal{E}}{\Omega_0}
\right],
\ee
where $\delta\mathcal{E}$ is the Zeeman splitting of the Landau levels, 
and $\Omega_0$ the cyclotron energy. They showed, that the $R_s$ 
depends on the orientation of the field re\-la\-tive not only to the conducting 
plane, but also to the staggered magnetization \mbox{(Fig. \ref{fig:real_space})}. 
Moreover, in a spin-flop configuration, where the staggered magnetization 
reorients itself transversely to the field, the Landau levels undergo \textit{no} Zeeman 
splitting \cite{revaz,zedr}, and the $R_s$ equals unity as long as the field ${\bf H}$ 
exceeds the spin-flop \mbox{threshold \cite{kabanov}.} This be\-ha\-vi\-or is in stark 
contrast to that of a two-dimensional pa\-ra\-mag\-ne\-tic conductor with isotropic 
Zeeman term $\mathcal{H}_Z = - \frac{1}{2} \mu_B g ({\bf H} \cdot {\bm \sigma})$, 
where the $R_s$ reads 
\be
\label{eq:R_s}
R_s = \cos \left[ \pi n \frac{g \mu_B m c}{\hbar e \cos \theta}\right], 
\ee  
with $\mu_B= \frac{1}{2} \frac{|e|\hbar}{m_e c}$ being the Bohr magneton, $m$ 
the cyclotron mass, and $\theta$ the inclination angle, sketched in the 
\mbox{Fig. \ref{fig:real_space}}: regardless of the value of $g$, the $R_s$ 
in the Eqn. (\ref{eq:R_s}) has infinitely many `spin-zeros' as a function of $\theta$. 

The peculiar behavior of the $R_s$, predicted in the Ref. \cite{kabanov}, stems 
from anisotropic spin-orbit character of the Zeeman coupling  $\mathcal{H}_Z$ 
in an antiferromagnet \cite{braluk,symshort,symlong}. 
The energy scale $E_{SO}$ of the relativistic spin-orbit coupling tends 
to be negligible compared with the antiferromagnetic gap $\Delta$ 
in the electron spectrum. Therefore, in a wide range of magnetic fields 
$E_{SO} \ll \langle \mathcal{H}_Z \rangle \ll \Delta$ con\-si\-de\-red hereafter, 
the Zeeman term is sensitive to the orientation of the field relative to the 
staggered magnetization, but not to the crystal axes. Hence, in this range 
of fields, the gyromagnetic factor $g$ in the Zeeman term 
turns into a tensor with two distinct eigenvalues, 
$g_\|$ and $g_\perp$, for the longitudinal (${\bf H}_\|$) and the transverse 
(${\bf H}_\perp$) components of the magnetic field ${\bf H}$ with respect 
to the staggered magnetization. 
The $g_\|$ is constant up to small relativistic corrections. By contrast, 
in $d$ dimensions, the $g_\perp$ must vanish on a $(d-1)$-dimensional 
manifold $\{\bf p^*\}$ in the Brillouin zone, due to a conspiracy of the crystal 
symmetry with that of the antiferromagnetic order \cite{symshort,symlong}.  
Thus, the $g_\perp$ must depend substantially 
on the quasiparticle momentum ${\bf p}$: 
\be
\label{eq:Zeeman_SOI}
\mathcal{H}_Z = - \frac{1}{2} \mu_B 
\left[
g_\| ({\bf H}_\| \cdot {\bm \sigma}) + 
g_\perp ({\bf p}) 
({\bf H}_\perp \cdot {\bm \sigma})
\right].
\ee 
Whenever a small 
carrier pock\-et is centered within the $\{\bf p^* \}$, the Zeeman splitting in 
a purely transverse field vanishes \cite{revaz,zedr,kabanov}, leading to a 
peculiar dependence of the $R_s$ on the field direction \cite{kabanov}. 
No spin-zeros appear beyond the spin-flop threshold, and such a behavior 
of the $R_s$ may serve as a signature of antiferromagnetic order. 
\begin{figure}[h]
\centerline{
   \mbox{\includegraphics[width=1.5in]{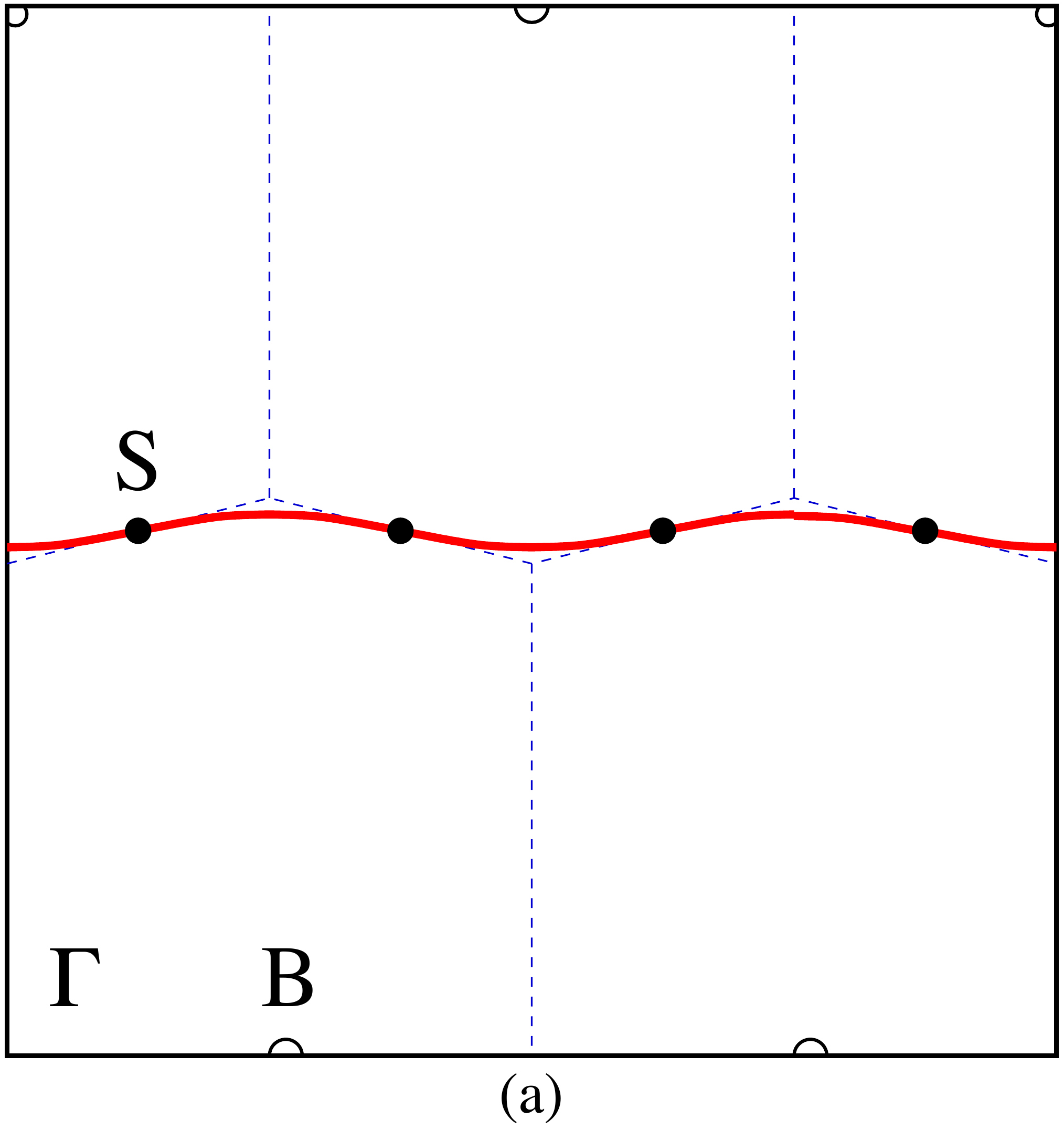}}
 \hspace{0.1cm}
   \mbox{\includegraphics[width=1.5in]{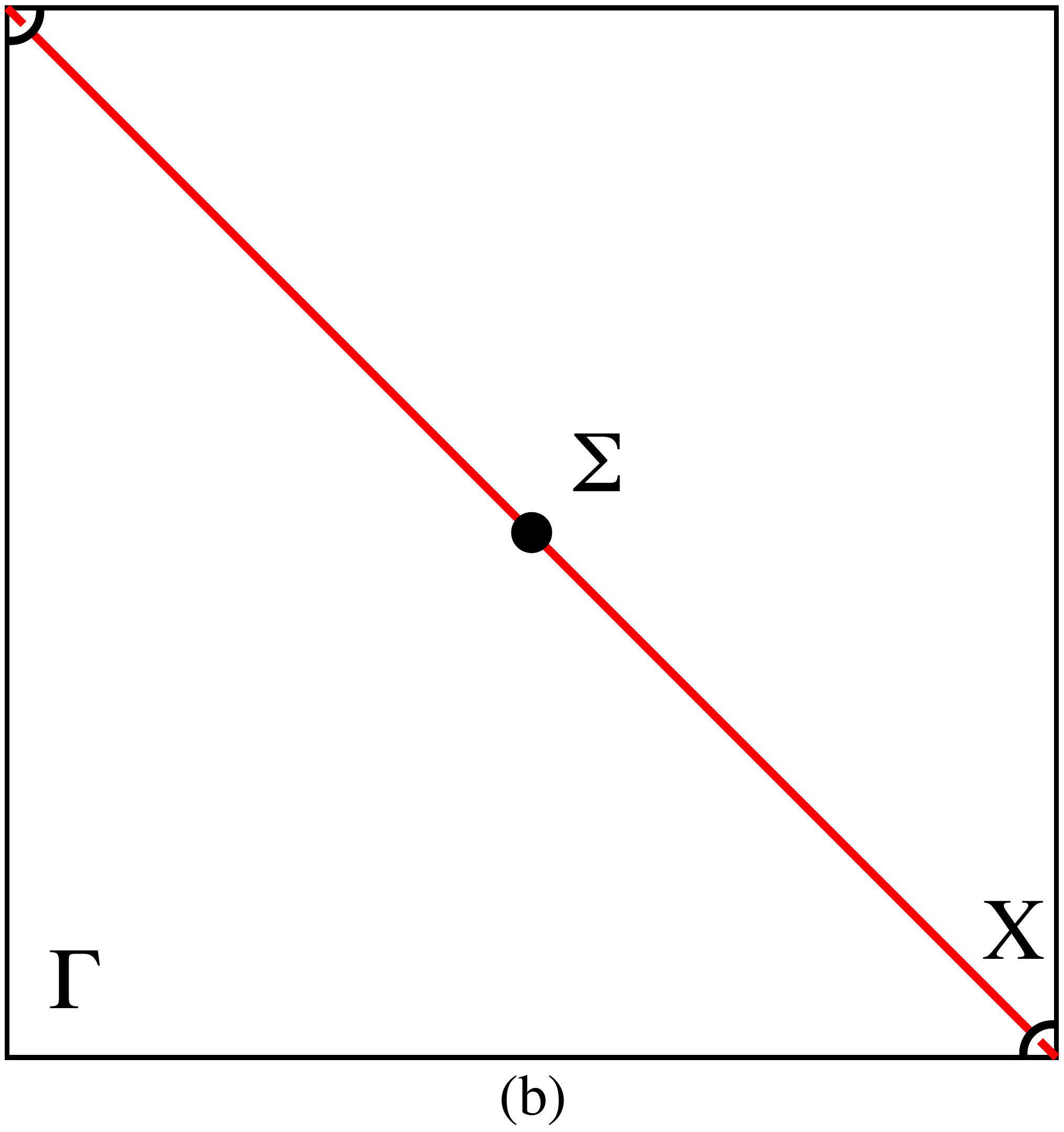}}
 \hspace{0.2cm}
 }
 \caption{(color online). 
(a) The first quadrant of the paramagnetic Brillouin zone of a 
\mbox{${\bf Q} = (\frac{3}{4}\frac{\pi}{a},\frac{\pi}{a})$} antiferromagnet 
\cite{millis}. The dashed (blue) lines denote the antiferromagnetic 
Brillouin zone boundaries. The thick (red) curve shows a typical line, 
where \mbox{$g_\perp ({\bf p}) = 0$}; this line is pinned by symmetry 
at the points $S$ at the momenta 
\mbox{${\bf p}^* = (\frac{\pi}{8a}[2n+1], \frac{\pi}{2a}[2 l +1])$}. The band 
extrema were found \cite{millis} at the points $B$, shown by the open 
circles, and, in a narrower parameter range, at the points $S$, shown 
by dark circles. 
(b)  The same, for a \mbox{${\bf Q} = (\frac{\pi}{a},\frac{\pi}{a})$} N\'eel 
antiferromagnet on a lattice of square symmetry. The thick (red) 
line shows the antiferromagnetic Brillouin zone boundary, where 
\mbox{$g_\perp ({\bf p}) = 0$}. The band extrema were found 
at the points $\Sigma$ (black circles) and $X$ (open circles).
}
\label{fig:BZ}
\end{figure}

A number of new developments suggest, that the antiferromagnetism 
in the underdoped YBCO may be weakly-incommensurate rather than 
commensurate, thus calling for an extension of the above results. 
Recent neutron scattering data \cite{haug} have shown evidence of 
incommensurate antiferromagnetism, induced by a magnetic field in the 
underdoped YBa$_2$Cu$_3$O$_{6.45}$ of very close composition to the 
samples of the Refs. 
\mbox{\cite{doiron,yelland,bangura,jaudet,sebastian_1,audouard}.}  
At the same time, a weakly-incommensurate stripe-like 
spin density wave with an ordering wave vector 
\mbox{${\bf Q} = \left(\frac{\pi}{a} \left[1 - \frac{1}{2N} \right], \frac{\pi}{a} \right)$} 
with an integer $N$ ($a$ being the lattice spacing) was found to yield 
\cite{millis,harrison}, in a broad parameter range, small electron 
pock\-ets, consistent not only with the quantum oscillation data 
\mbox{\cite{doiron,yelland,bangura,jaudet,sebastian_1,audouard},} 
but also with the observed negative low-temperature Hall coefficient \cite{leboeuf}. 

How could such a weakly-incommensurate  antiferromagnetism manifest itself 
in quantum oscillations? The answer depends on the location of the carrier 
pock\-et in the Brillouin zone. Pock\-ets, centered within the $\{ {\bf p^*} \}$, 
were described above. Weak in\-com\-men\-su\-ra\-bi\-li\-ty opens 
a new possibility: pock\-ets, centered outside the $\{ {\bf p^*} \}$. 

For  \mbox{${\bf Q} = \left(\frac{3}{4} \frac{\pi}{a} , \frac{\pi}{a} \right)$} 
and generic values of the density wave parameters, the Ref. \cite{millis} 
found such pock\-ets, centered at the points $B$ in the Fig. \ref{fig:BZ}(a), 
while the Ref. \cite{harrison} found analogous pockets for 
\mbox{${\bf Q} = \left(\frac{7}{8} \frac{\pi}{a} , \frac{\pi}{a} \right)$}. 
These pockets are about $\frac{\pi}{2a}$ away from the nearest point $S$, 
where the line $g_\perp ({\bf p}) = 0$ is pinned by symmetry. In the simplest 
case, the line $g_\perp ({\bf p}) = 0$ is singly-connected and pinned at the 
points $S$; the $g_\perp ({\bf p})$ is suppressed only within momentum 
deviations $| \delta p | \lesssim \xi^{-1} \ll \frac{\pi}{2a}$ from this  line 
\cite{symlong}. In such a case, the $g$-tensor at the $B$-pockets is isotropic 
up to va\-ni\-shing\-ly small corrections of the order of $( a / \xi )^2 \ll 1$, 
which can be read off the Eqn. (11) of the Ref. \cite{symlong} for the 
\mbox{${\bf Q} = \left(\frac{\pi}{a}, \frac{\pi}{a} \right)$} N\'eel order. 

However, a very recent study \cite{norman} found the $g_\perp ({\bf p}) = 0$ 
line numerically for a \mbox{${\bf Q} = \left(\frac{3}{4} \frac{\pi}{a} , \frac{\pi}{a} \right)$} 
spin density wave, and discovered that this line may be multiply-connected, with 
components, disconnected from symmetry-enforced degeneracy points. 
Some of these components were found to pass close to the $B$-points. 
In such cases, the $g_\perp ({\bf p})$ 
for the $B$-pockets is non-zero yet reduced, and thus the $g$-tensor is 
strongly anisotropic \cite{norman}. By contrast with the pockets, centered 
on the line $g_\perp ({\bf p}) = 0$, the Zeeman splitting of the $B$-pocket 
Landau levels does {\em not} vanish, and the spin-zeros do appear even 
in the spin-flop configuration, albeit at greater inclination angles $\theta$.

Do the above observations open any diagnostic opportunities? 
Of course, spin-zeros are no proof of antiferromagnetism. 
However, having experimental knowledge of the presence and periodicity 
of the antiferromagnetism in the sample greatly restricts the allowed 
possibilities: for instance, 
in \mbox{${\bf Q} = \left(\frac{3}{4} \frac{\pi}{a}, \frac{\pi}{a} \right)$} 
and \mbox{${\bf Q} = \left(\frac{7}{8} \frac{\pi}{a}, \frac{\pi}{a} \right)$}  
spin density wave states, the $B$ points in the Fig. \ref{fig:BZ}(a) 
were the \textit{only} band extrema outside the $\{ {\bf p^*} \}$, found by the 
Refs. \cite{millis,harrison} for generic parameter values. Thus, observation 
of spin-zeros in such an antiferromagnet constrains the detected carrier 
pocket uniquely to the center point $B$ of the magnetic Brillouin zone.  
 
By contrast, in a 
\mbox{${\bf Q} = \left(\frac{\pi}{a}, \frac{\pi}{a} \right)$} antiferromagnet, 
in the relevant parameter range the calculated band mi\-ni\-ma were 
found \textit{only} on the magnetic Brillouin zone boundary \cite{chen}, 
where $g_\perp({\bf p}) = 0$.  For such carrier pockets, no spin-zeros 
appear in a purely transverse field; thus, observation of spin-zeros 
is essentially incompatible with 
\mbox{${\bf Q} = \left(\frac{\pi}{a}, \frac{\pi}{a} \right)$} 
N\'eel antiferromagnetism.

The experiments have not yet reached a consensus. Measurements 
of the underdoped YBa$_2$Cu$_3$O$_{6.54}$ have found no spin-zeros  
within the expected angular range \cite{sebastian_2}. By contrast, the 
Ref. \cite{ramshaw} studied the underdoped YBa$_2$Cu$_3$O$_{6.59}$, and did 
find spin-zeros, consistent with isotropic $g$-tensor, within the range of the Ref. 
\cite{sebastian_2}.

While settling this disagreement is beyond the scope of the present work, 
eventually finding \textit{no} spin-zeros at all would be consistent with 
antiferromagnetism and the pockets centered within the $\{ {\bf p^*} \}$. 
By contrast, between the 
\mbox{${\bf Q} = \left(\frac{\pi}{a}, \frac{\pi}{a} \right)$} and 
\mbox{${\bf Q} = \left(\frac{\pi}{a} \left[1 - \frac{1}{2N} \right], \frac{\pi}{a} \right)$}  
spin density waves, detecting spin-zeros would be consistent only with the 
latter periodicity and with the detected pockets centered uniquely at the $B$ 
points in the Fig. \ref{fig:BZ}(a). 
                                                        
I will now demonstrate the symmetry underpinnings of the above results \cite{symlong}. 
In a \mbox{${\bf Q} = \left(\frac{\pi}{a} \left[1 - \frac{1}{2N} \right], \frac{\pi}{a} \right)$} 
spin density wave state with an integer $N$ and possible charge modulations at 
multiples of the ${\bf Q}$, the conduction electron spin ${\bm \sigma}$ is subject 
to the exchange coupling ${\bf \Delta (r)} \cdot {\bm \sigma}$, changing sign upon  
translation ${\bf T_b}$ by a single lattice spacing along the $y$ axis, or by $2N$ 
spacings along the $x$ axis: ${\bf \Delta (r+b)} = - {\bf \Delta (r)}$. 
Hence, in a transverse magnetic field, $\theta {\bf T_b U_n}(\pi)$ is an anti-unitary 
symmetry of the Hamiltonian, where $\theta$ is time reversal, and ${\bf U_n}(\pi)$ 
is a spin rotation by $\pi$ around the unit vector ${\bf n}$ of the staggered 
magnetization. Retracing the derivation of the Eqn. (5) in the Ref. \cite{symlong}, 
one finds
\be 
\label{eq:thetaTU}
\langle 
{\bf p}
 | 
\theta {\bf T_b U_n}(\pi)
 | {\bf p}  
\rangle 
 = 
e^{-2 i {\bf p \cdot b}}
\langle 
{\bf p}
 | 
\theta {\bf T_b U_n}(\pi) 
 | {\bf p}  
\rangle.
\ee 
Thus, a Bloch eigenstate $| {\bf p} \rangle$ at a momentum ${\bf p}$ 
is or\-tho\-go\-nal to its partner $\theta {\bf T_b U_n}(\pi) | {\bf p} \rangle$ 
at the momentum $-{\bf p}$ \cite{footnote_2}. In the folded Brillouin zone, 
defined by the periodicity of the ${\bf \Delta (r)}$, the momenta 
\mbox{${\bf p}^* = (\frac{\pi}{2Na}[2k + 1], \frac{\pi}{2a})$} 
and $-{\bf p}^*$ are equivalent for an integer $k$. Hence, the Eqn. (\ref{eq:thetaTU}) 
proves the Kramers degeneracy of the Bloch eigenstates at ${\bf p = p}^*$ 
in a transverse magnetic field. In two dimensions, the equation 
$g_\perp ({\bf p}) = 0$ defines a line in the Brillouin zone, and the 
Eqn. (\ref{eq:thetaTU}) pins this line at the above symmetry-enforced 
degeneracy points $S$, as shown in the Fig. \ref{fig:BZ}(a) 
for \mbox{${\bf Q} = \left(\frac{3\pi}{4a}, \frac{\pi}{a} \right)$}. 

The $S$ points do tend to host a band extremum \cite{millis,harrison}.
The leading term of the momentum expansion of the $g_\perp ({\bf p})$ around 
these points is linear, and the Landau levels and their Zeeman splitting have 
been described in the Refs. \cite{revaz,kabanov,zedr}.  
A carrier pocket may also be centered at a point, where the line 
$g_\perp ({\bf p}) = 0$ intersects itself, as it does at the point $X$ in the 
Fig. \ref{fig:BZ}(b). The leading term of the momentum expansion of the 
$g_\perp ({\bf p})$ around the point $X$ is quadratic \cite{revaz,symlong}, 
and the carrier Hamiltonian near the point $X$ takes the form 
\be
\label{eq:Hamiltonian_X}
\mathcal{H} = \frac{{\bf p}^2}{2m} - 
({\bf \Omega}_\| \cdot {\bm \sigma}) - 
\frac{p_x^2 - p_y^2}{2m \Delta} 
({\bf \Omega}_\perp \cdot {\bm \sigma}),
\ee 
where ${\bf \Omega} \equiv \frac{1}{2} g_\| \mu_B {\bf H}$.  
The small pocket size implies, that 
\mbox{$\frac{p_F^2}{m \Delta} \sim \frac{\mu}{\Delta} \ll 1$}, 
where $\mu$ is the chemical potential, counted 
from the bottom of the pocket.

\begin{figure}[h]
 \hspace{3cm}
 \epsfxsize=8cm
 \includegraphics[width=3in]{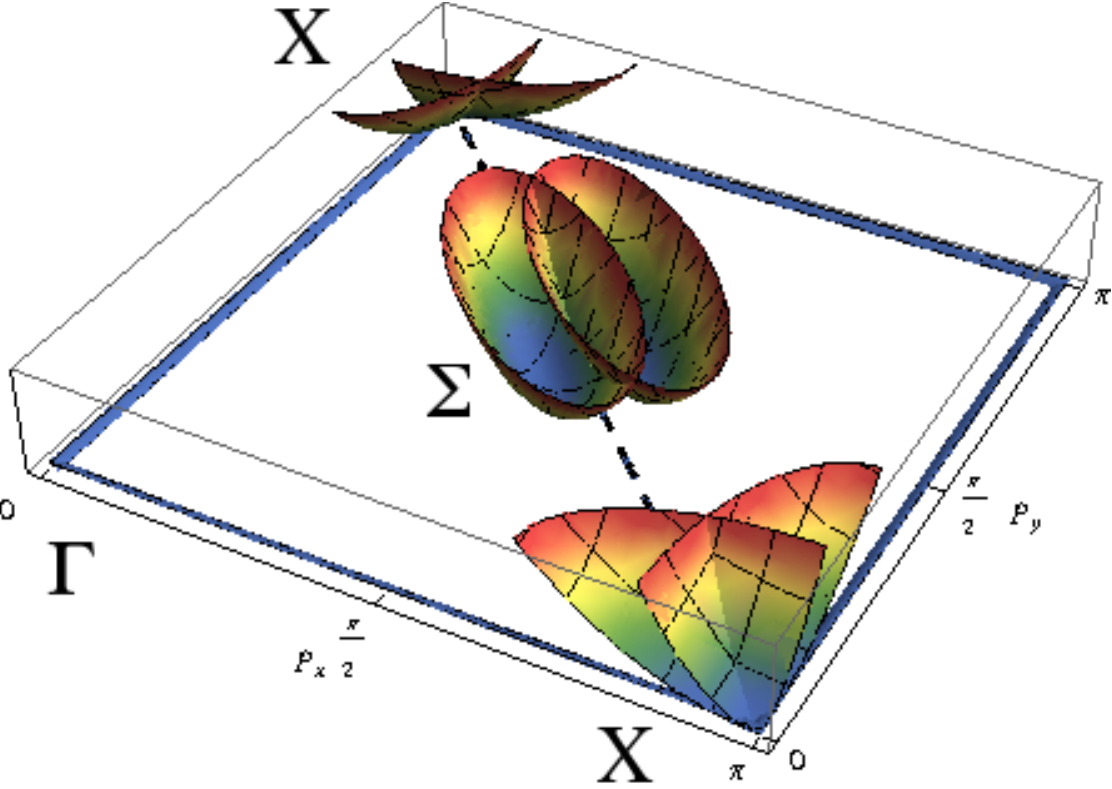}
 \vspace{15pt}
\caption{(color online). 
A sketch of the Zeeman splitting of the small carrier pockets, centered 
at the points $X=(0,\frac{\pi}{a})$ [Eqn. (\ref{eq:Hamiltonian_X})]  and 
$\Sigma=(\frac{\pi}{2a},\frac{\pi}{2a})$ [Eqn. (\ref{eq:Hamiltonian_Sigma})] 
in the first quadrant of the Brillouin zone, in a purely transverse magnetic 
field.  The dashed line, passing through the points $X$ and $\Sigma$, 
is the magnetic Brillouin zone boundary, where $g_\perp({\bf p}) = 0$. 
The pocket sizes and the splitting are greatly exaggerated. 
} 
\label{fig:zone_splitting}
\end{figure}

According to the Hamiltonian (\ref{eq:Hamiltonian_X}), 
in a transverse field (${\bf \Omega}_\|=0$) the Landau levels 
undergo no Zeeman splitting, while the effective mass tensor 
becomes anisotropic and dependent on the spin projection 
onto ${\bf \Omega}_\perp$ as per 
\mbox{$m_{x/y}^{-1} = m^{-1} \left[1 \pm 
\frac{({\bf \Omega}_\perp \cdot {\bm \sigma})}{\Delta}\right]$}, 
as shown in the Fig. \ref{fig:zone_splitting}. 
Beyond the spin-flop threshold, the staggered magnetization 
re-orients itself transversely to the field; 
thus, the Landau levels undergo no Zeeman splitting,
and no spin-zeros are to be found at any field direction. 

Near spin-flop but with ${\bf \Omega}_\| \neq 0$ in the 
\mbox{Hamiltonian (\ref{eq:Hamiltonian_X}),} the Zeeman splitting 
$\delta \mathcal{E}$ of the Landau levels is simply 
\mbox{$\delta \mathcal{E} = 2 \Omega_\|$} \cite{revaz}, with small 
corrections of the order of $[\mu/\Delta]^2 \ll 1$:  at a low enough 
doping, $\delta \mathcal{E}$ behaves as if the last term in 
the Eqn. (\ref{eq:Hamiltonian_X}) simply vanished.  

Hence, according to the Eqn. (\ref{eq:amplitude}), for a small 
pocket at the point $X$, the field direction of the $l$-th 
spin-zero in the main harmonic ($k=1$) satisfies the equation 
\be
\label{eq:spin_zeros_X}
\frac{\delta\mathcal{E}}{\Omega_0} = 
\eta \frac{H_\|}{H_0}
 = 
 \eta
 \cdot 
   \tan \theta \cdot \cos \varphi 
  = l + \frac{1}{2}, 
\ee
where $\eta = g_\| \mu_B \frac{mc }{\hbar e} = \frac{g_\|}{2} \frac{m}{m_0}$, 
$l$ is an integer, 
\mbox{$H_0 = H \cos \theta$} is defined in the Fig. \ref{fig:real_space}, 
and \mbox{$H_\| = H \sin \theta \cos \varphi$} is the longitudinal 
component of the field with respect to the staggered magnetization. 

The distinction between the above spin-zeros and those of the $S$- and 
$\Sigma$-pockets stems from the leading term of the momentum expansion 
of the $g_\perp ({\bf p})$ around the points $S$ and $\Sigma$ being linear 
rather than quadratic: 
\be
\label{eq:Hamiltonian_Sigma}
\mathcal{H} = \frac{p_x^2}{2m_x} + \frac{p_y^2}{2m_y}
 - ({\bf \Omega}_\| \cdot {\bm \sigma}) - \frac{\xi p_y}{\hbar}
({\bf \Omega}_\perp \cdot {\bm \sigma}), 
\ee 
where $p_y$ is the transverse component of the momentum with respect 
to the magnetic Brillouin zone boundary in the Fig. \ref{fig:zone_splitting}. 
Here, as at the point $X$, the carrier pock\-et is assumed small enough 
to be described by the \mbox{Eqn. (\ref{eq:Hamiltonian_Sigma}):}  
\mbox{$\frac{\xi p_y}{\hbar} \lesssim \sqrt{\frac{\mu}{\epsilon^*}} \ll 1$}, where 
\mbox{$\epsilon^* = \frac{\hbar^2}{2 m_y \xi^2} \sim \frac{\Delta^2}{\epsilon_F}$}, 
and $\mu$ is the chemical potential, counted from the bottom of the pock\-et. 
The length scale $\xi$ is of the order of the antiferromagnetic co\-he\-rence 
length $\hbar v_F / \Delta$ \cite{symlong}. The spin-zeros for such a pock\-et, 
encapsulated in the Eqn. (11) of the Ref. \cite{kabanov}, differ from those 
given by the Eqn. (\ref{eq:spin_zeros_X}) only via the small parameter  
\mbox{$\sqrt{\frac{\mu}{\epsilon^*}} \ll 1$}. This quantitative and, for most field 
orientations, numerically small difference is likely to render experimentally 
distinguishing the $\Sigma$ pockets from their $X$ counterparts rather difficult, 
especially on the background of the Fermi surface corrugation \cite{harrison_2} 
and bilayer splitting \cite{audouard}. These effects also modify the 
oscillation amplitude in a material-specific way \cite{audouard,harrison_2}. 

To conclude, I have shown that, in an antiferromagnet, a combination of symmetry  
arguments with the knowledge of the possible positions of the band extrema 
 \cite{footnote_3} allows either to constrain the possible locations of a small carrier 
 pock\-et, or even to pinpoint it in the Brillouin zone by mapping the spin-zeros 
 of the quantum oscillation amplitude. This opportunity arises due to the anisotropic 
 spin-orbit character of the Zeeman coupling in an antiferromagnet, 
 and does not exist in a paramagnetic conductor. While I use the 
\mbox{${\bf Q} = \left(\frac{\pi}{a} \left[1 - \frac{1}{2N} \right], \frac{\pi}{a} \right)$} 
and 
\mbox{${\bf Q} = \left(\frac{\pi}{a}, \frac{\pi}{a} \right)$} 
spin density waves as an illustration, possibly re\-le\-vant to cuprate superconductors, 
the method is applicable to many other antiferromagnets such as iron pnictides, 
organic and heavy fermion materials.

I am grateful to the Condensed Matter Theory groups of the ICTP in Trieste and 
MPI PKS in Dresden, where parts of this work were done, for the kind hospitality. 
I thank C. Capan, B. Vignolle, D. Vignolles, and especially C. Proust for the helpful  
comments, and M. R. Norman and A. J. Millis for extended discussions.


\begin{thebibliography}{99}
\bibitem{footnote_2} 
On the lines $p_x = k \frac{\pi}{a}$ and $p_y = k \frac{\pi}{Na}$ 
with $k = 0, \pm 1, \pm 2, ...$, the exponent in the r.h.s. of the 
Eqn. (\ref{eq:thetaTU}) equals unity, and the degeneracy is not 
protected.
\bibitem{footnote_3} 
The band extrema tend to fall at the high-symmetry points 
in the Brillouin zone, and are readily identified by re\-la\-ti\-ve\-ly simple 
calculations such as those cited above.
\end{thebibliography}
\end{document}